\documentclass[times,12pt]{article}

\usepackage[normalem]{ulem}
\usepackage{amssymb}
\usepackage{amsmath}
\usepackage{amsfonts}
\usepackage{amsthm}
\usepackage{graphicx}
\usepackage{authblk}
\usepackage{subfigure}
\usepackage{framed}
\usepackage{color}
\usepackage[toc,page]{appendix}

\usepackage[authoryear]{natbib}
\usepackage{url}

\numberwithin{equation}{section}

\newcommand{\s}{\mathbf{s}} 
\newcommand{\x}{\mathbf{x}}

\newcommand{\N}{\mathcal{N}}

\newcommand{\sigmasK}{\mathbf{\Sigma}_{\mathbf{s} \setminus L}}

\newcommand{\sK}{\mathbf{s}_{\setminus L} }

\newcommand{\C}{\mathbf{C}} 
\newcommand{\hC}{\hat C} 
\newcommand{\tC}{\mathbf{C }_L}
\newcommand{\tS}{\mathbf{\Sigma }_{\mathbf{s}, L}^\eta}  
\newcommand{\Sx}{\mathbf{\Sigma }_{\mathbf{x} }}
\newcommand{\Ss}{\mathbf{\Sigma }_{\mathbf{s}}}

\newcommand{\tvj}{v_L}

\newcommand{\gamk}{\gamma_L}
  
 \newcommand{\bs}{\bar s} 
\newcommand{\bx}{\bar x} 
\newcommand{\cL}{\mathcal L}

\newcommand{\sigx}{\sigma_x^2} 

\newcommand{\sigs}{\sigma_s^2}

\newcommand{\Nchoosen}{\begin{pmatrix} N \\ n \end{pmatrix}}

\newcommand{\dTD}{d_{\text{TD}}}

\newcommand{\0}{\mathbf 0} 
\newcommand{\rs}{\rho_s} 
\newcommand{\rx}{\rho_x}

\title{\textbf{Visual decisions in the presence of measurement and stimulus correlations}}
\author[1]{Manisha Bhardwaj \thanks{manisha@math.uh.edu}}
\author[1]{Sam Carroll  \thanks{srcarroll314@gmail.com}}
\author[2,3]{Wei Ji Ma \thanks{weijima@nyu.edu}}
\author[1,4]{Kre\v{s}imir Josi\'{c} \thanks{josic@math.uh.edu}}
\affil[1]{Department of Mathematics, University of Houston}
\affil[2]{Department of Neuroscience, Baylor College of Medicine}
\affil[3]{Now at: Center for Neural Science and Department of Psychology, New York University}
\affil[4]{Department of Biology and Biochemistry, University of Houston}
\date{}

\begin{document}

\maketitle

\begin{abstract}
Humans and other animals base their decisions on noisy sensory input.  Much work has 
therefore been devoted to understanding the computations that underly such decisions. 
The problem has been studied in a variety of tasks and with stimuli of differing complexity.  
However, the impact of correlations in sensory noise on perceptual judgments is not well understood.
Here we examine how stimulus correlations together with correlations in sensory noise impact
decision making.  As an example, we consider the task of detecting the presence of a single or multiple targets amongst 
distractors.  We assume that both the distractors and the observer's measurements of the stimuli
are correlated.  The computations of an optimal observer in this task are nontrivial, yet can be analyzed and understood intuitively.
 We find that when distractors are strongly correlated, measurement correlations can have 
 a strong impact on performance.  When distractor correlations are weak, measurement correlations have
little impact, unless the number of stimuli is large.  Correlations in neural responses to structured stimuli 
can therefore strongly impact perceptual judgments.
\end{abstract}

\section{Introduction}

The perceptual system has evolved to extract ecologically meaningful information from sensory input.  For example, in many mid- to high-level visual tasks the brain has to make categorical, global judgements 
based on multiple stimuli where the identity of any individual stimulus is not of direct relevance.  In a visual search task, the goal might be to detect whether a predefined target object is present in a scene that contains multiple objects.  Complicating such tasks is the fact that 
noise corrupts sensory measurements, especially when 
observation time is short, or many objects are present.

Much work has been devoted to modeling the decision processes by which the brain converts  noisy sensory measurements of a set of stimuli into a judgement about a global world state, such as the presence or absence of a target. These models often focus on various decision rules that can be applied to the measurements. By contrast, the measurements themselves are usually modeled in a rather stereotypical fashion, namely as independent and normally distributed, (e.g.~\cite{peterson1954,nolte1967,pelli1985,graham1987,palmer1993,baldassi2000,baldassi2002,VandenBerg2012,Ma2011,Mazyar2012}). Both the assumption of independence and the assumption of Gaussianity can be questioned. Specifically, neural correlations can extend to distances as long as 4mm in monkey cortex~\citep{Ecker:2010dna,Cohen:2011}.  This suggests that  sensory measurements can be strongly correlated~\citep{Rosenbaum:2010eq,Chen:2006}. Here we focus on the effects of violation of the assumption of independent measurements on performance in categorical, global perceptual judgements.  

To make such perceptual judgments, an observer needs to take into account  
the statistical structure of  the stimuli \emph{and} the structure of measurements. Consider a search task where a subject is required to detect a target 
among distractors.  
The effects of measurement correlations and stimulus correlations will be intertwined:  
If the distractors are identical on a given trial, then strong correlations between the measurements will help preserve their perceived similarity. Namely, an observer can group the  distractor measurements and identify the target as corresponding to the outlying measurement.  By contrast,  when distractors are unstructured (independently drawn across locations), strong measurement correlations may have no effect on performance.  Thus, 
measurement and stimulus correlations should  not be considered in isolation.

Here we examine how measurement and stimulus correlations impact the strategy and the performance
of an ideal observer in a target detection task.  We assume that on half the
trials, one or more target stimuli are presented along with a number of distractors, whereas on the other half of trials, only distractors are presented.
The task is to infer whether targets are present or not.  Importantly, we assume that
the distractor stimuli are not drawn independently -- for instance, in the extreme case the targets could be identical.  
Our ideal observer infers target presence based on measurements of
the stimuli.  We assume that these measurements are corrupted by \emph{correlated}
noise.  In an extreme case, this noise is perfectly correlated and all measurements are
perturbed by the same, random value. 

We provide an analytical study of the optimal decision rule to 
show that the interplay of measurement and stimulus correlations can be intricate.
In general, if the stimuli are strongly correlated, then measurement correlations can 
strongly affect the performance of an ideal observer.  When stimuli are weakly correlated,
measurement correlations have a smaller impact.  

We expect that these insights hold more generally:  Natural stimuli are structured, and their distributions
are concentrated along low dimensional structures in stimulus space~\citep{Geisler:2008}.  Correlations in measurement
noise could thus help in the inferring parameters of interest. 
Ultimately, we thus expect  our results to be relevant to modeling perceptual decision-making in natural scenes.

\section{Model description}

To examine how decisions of an ideal observer are determined by the statistical structure of measurements and stimuli we consider the following task.  An observer is asked whether  target stimuli are present among a set of distractor stimuli. Each stimulus, $i$,  is characterized by a scalar, $s_i \in {\mathbb R}$. The set of $N$ stimuli presented on a single trial is  characterized by the vector $\s = (s_1, s_2, \cdots, s_N)$.  For instance, stimuli could be pure tones characterized by their frequency,  gratings characterized by their orientation, or 
ellipses characterized by the orientation of their major axis.
A \emph{target} is a stimulus with a particular characteristic, $s_T$.  A target could be a vertical grating, or a pure tone at 440Hz. For simplicity, we assume that, if $i$ is a target, then 
$s_i = s_T = 0$, and stimulus 
characteristics are measured relative to that of a target.    Stimuli that are not targets are distractors. For such stimuli $s_j \neq 0$ with probability~1. We will consider situations with single and multiple targets.

\begin{figure}[htp!]
\begin{center}
\includegraphics[scale = 0.4]{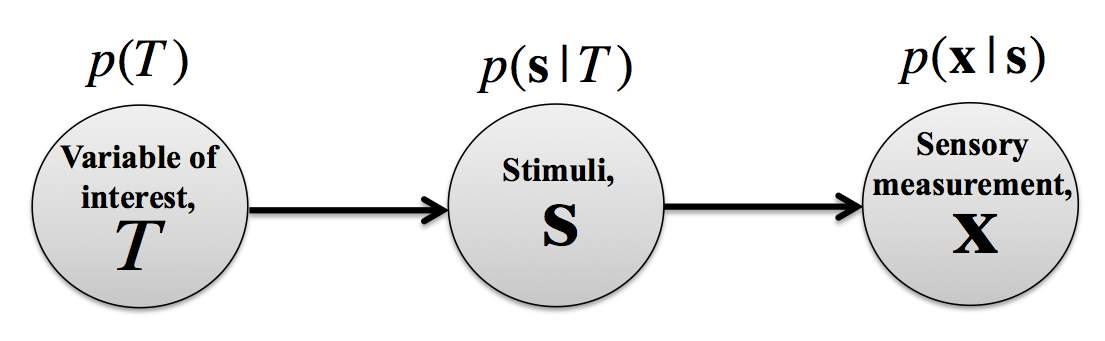}
\caption{A schematic representation of  Bayesian inference. Information about a parameter of interest, $T$, is encoded in a stimulus vector, $\mathbf{s}$.  The dimension of $\mathbf{s}$ could be much higher than the dimension of $T$.  The observer makes a sensory measurement, $\mathbf{x}$, of the stimulus, $\mathbf{s}$, and must infer $T$ from this measurement.  This inference requires marginalization over $\mathbf{s}$. }
\end{center}
\vspace{-5mm}
\label{fig:intro_marginalization}
\end{figure}

The following treatment parallels the one we used earlier for independent measurements~\citep{Mazyar2012, Mazyar2013, Bhardwaj2015}. We denote target presence by $T = 1$, and absence by $T  = 0$.  We assume that targets are present with probability 0.5.  
When $T=0$ (no targets) stimuli are drawn from a multivariate normal distribution with mean $\0_N = (0, 0, \cdots, 0)$, and covariance matrix, $\Ss$.  The subscript denotes vector length, so  $\0_N$ has $N$ components.  We can therefore write
\begin{equation}\label{eq:dist_s}
p(\s | T) =  \N(\s; \0_N, \Ss), 
\end{equation}
where $\N(\s; \boldsymbol{\mu}, \boldsymbol \Sigma)$ denotes the density of the normal distribution with mean $\boldsymbol{\mu}$ and covariance $\boldsymbol \Sigma$.
For simplicity, we assume $\Ss$ has constant diagonal and off-diagonal terms, so that
\begin{equation}\label{eq:matrix_Sigmas}
\Ss = \sigs 
\begin{bmatrix} 
1 & \rs  & \cdots & \rs \\
\rs  & 1 & \cdots & \rs  \\
\vdots & & \ddots & \vdots \\
\rs  & \rs & \cdots &1
\end{bmatrix}.
\end{equation}
The correlation coefficient, $\rs$, determines the relation between the components of the stimulus.

If $T=1$, then $n\geq 1$ targets are present. In this case we assign, with uniform probability, the target characteristic to $n$ out of the total of $N$ stimuli. This subset $n$ targets is denoted by $L$.  We denote by $\cL$ the collection of all $\Nchoosen$ possible choices of the sets $L$ of  targets. 

The remaining  $N-n$ distractors are drawn from a multivariate normal distribution with mean $\0_{N-n}$, and covariance matrix $\sigmasK$ of dimension $(N-n) \times (N-n)$.  Let $\s_L = (s_{i_1}, s_{i_2},\cdots,s_{i_n}), \; i_l \in L$ denote the  target stimuli, and $\sK = (s_{j_1}, s_{j_2}, \cdots, s_{j_{N-n}}), \; j_l \notin L$ the distractors.  We can therefore write 
\begin{equation*}
p(\s_L |T = 1) = \sum_{i \in L} \delta(s_i),  \qquad \text{and} \qquad
p(\sK |T = 1) = \N(\sK; \0_{N-n}, \sigmasK).
\end{equation*}  
Since the density $p(\s | T = 1, L)$ is singular, we introduce the auxiliary covariance,
\begin{equation} \label{E:coveta}
(\tS)_{i,j} = 
\begin{cases}
(\sigmasK)_{i,j}, & \text{ if } i, j \notin L, \\
\eta, & \text{ if } i = j \in L, \\ 
0, & \text{ if } i \in L, \text{ or } j \in L, \text{ and }  i \neq j,
\end{cases}
\end{equation} 
where $\eta > 0$. Therefore,
$$p(\s | T = 1, L) = \lim_{\eta \downarrow 0} \; \N(\s;\0_N, \tS).$$ 

We further assume that an observer makes a noisy measurement, $x_i$,  of each stimulus, $s_i$.  This measurement can be thought of as the estimate of stimulus $i$ obtained from the activity of a population of neurons that responded to the stimulus. We denote by $\x = (x_1, x_2, \cdots, x_N)$ the vector of $N$ measurements.   
It is commonly assumed that these measurements are unbiased, and corrupted by additive, independent, normally distributed noise, so that
$
p(\x | \s) = \prod_{i=1}^n \N(x_i;s_i, \sigma_x^2).
$
Here we consider the more general situation where the measurements are unbiased, but noise could be correlated so that
\begin{equation}\label{eq:dist_x}
p(\x | \s) =   \N(\x; \s, \Sx). 
\end{equation} 
We consider the particular case when $\Sx$ has constant diagonal terms, $\sigx,$ and off-diagonal terms, $\rx \sigx$, so that
\begin{equation}\label{eq:matrix_Sigmax}
\Sx = \sigx
\begin{bmatrix} 
1 & \rx  & \cdots & \rx \\
\rx  & 1 & \cdots & \rx  \\
\vdots & & \ddots & \vdots \\
\rx  & \rx & \cdots & 1
\end{bmatrix}.
\end{equation}

%

\begin{figure}[htp!]
\begin{center} 
\includegraphics[scale = 1]{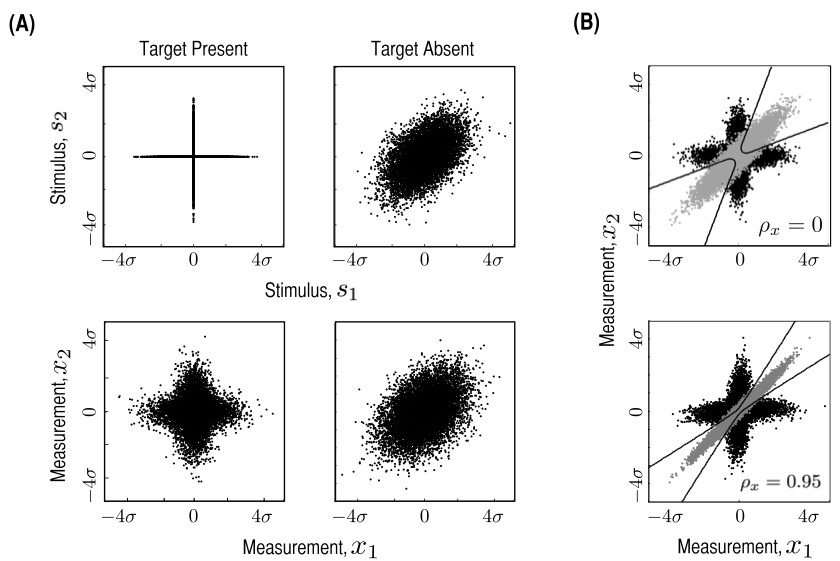}
\caption{
Stimulus and measurement distributions in the single target detection task with $N = 2$ stimuli and $\sigma_s = 15, \sigma_x = 4$. \textbf{(A)} (Top) Stimulus distributions on target present (left) and target absent (right) trials for $\rs = 0.5$. On target present trials, the distribution is constrained to the axes, since one of the stimuli is a target. On target absent trials, the two stimuli follow a bivariate normal distribution. (Bottom) Distributions of measurements in the absence of measurement correlations, $\rho_x = 0$.  On both the target present  (left), and target absent case (right), the measurement distribution inherits its shape from the stimulus distribution. \textbf{(B)} Overlap of measurement distributions for strongly correlated distractors, $\rs = 0.99$. Measurement distributions on target present (black) and absent (gray) trials are shown for $\rx = 0$ (top) and $\rx = 0.95$ (bottom). The overlap between the measurement distributions decreases with increasing measurement correlation, $\rx$. The  decision boundary (solid line) therefore better separates the two distribution when $\rs$ and $\rx$ are both large.   We used high correlations to bring out the difference between the distributions. Axes represent the total standard deviation of the measurements,  $\sigma = \sqrt{\sigs + \sigx}$. }
\end{center}
\label{F:overlap}
\end{figure}

%

Note that
$$
p(\x | T = 1) =   \int p(\x | \s) p(\s | T = 1) d\s,
$$
and similarly for $p(\x | T = 0)$.  We must thus marginalize over $\s$ to obtain these distributions.
A simple computation (See Appendix~\ref{appendix_multipleTST_eq.d1}) shows that 
$p(\x | L, T=1) = \N(\x; \0_N, \mathbf{\Sigma }_{\mathbf{s}, L}^0 + \Sx)$, and 
$p(\x | T=0) = \N(\x; \0_N, \Ss + \Sx)$.

\section{Results}
\label{section:Results}

Our goal is to describe how correlations between stimuli along with correlations between their measurements 
 affect the decisions of an optimal observer in a target detection task.   
We examine how performance changes as both correlations between distractors and between measurements are 
varied.    

The stimuli, $\s$, follow different distributions 
depending on whether $T=0$ or $T=1$.  Measurement noise increases
the overlap between the corresponding measurement distributions, 
$p(\x | T=0)$ and $p(\x | T=1)$.  The higher the overlap between these two distributions, 
the more difficult it is to tell whether a target is present or not.  However, correlations in measurement noise can reduce such overlap (See Fig.~\ref{F:overlap}B), even when noise intensity  is unchanged.  Therefore, the estimate of a parameter from a neural response depends not only on the level, $\sigma_x$, but also on the structure of measurement noise~\citep{Abbott1999,Sompolinsky2001,Averbeck2006,Josic2009}.

 An ideal observer makes a decision based on the sign of the log posterior ratio,
\begin{equation}\label{eq:decision_TD}
\dTD(\x) = \log \frac{p(T = 1| \x)}{p(T = 0| \x)}  = \log \frac{p(\x | T = 1)}{p(\x | T = 0)} + \log \frac{p(T = 1)}{p(T = 0)}.
\end{equation}  
If $\dTD(\x) > 0$, the observer infers that a target is present.   Note that 
$$ 
p(\x | T = 1) = \sum_{L \in \cL} p(\x | L, T = 1) p(L). 
$$
Given that the prior probability that  $L$ is a set of targets is uniform gives $p(L) =\Nchoosen^{-1}$, and
\begin{equation} \label{E:targets}
\dTD(\x) = \log \frac{p(\x | T = 1)}{p(\x | T = 0)} = \log \left[ \Nchoosen^{-1} \sum_{L \in \cL}\frac{ p(\x | L, T = 1)}{p(\x | T = 0)} \right].
\end{equation}
The decision variable thus depends on the sum of the normalized probabilities that a measurement in
$\x$ is made, given that the target set is $L$.

Note also that $$p(L | \x, T = 1) \propto p(\x | L, T=1).$$  Therefore, the decision variable can also 
be interpreted as a sum of likelihoods that $L$ is a target set, given a measurement $\x$. Thus the
decision is directly related to the posterior distribution over the target sets $L$:  The summands  in Eq.~\eqref{E:targets} correspond to the evidence that  $L$ is a set of targets.

The distribution of measurements, $\x$, depends on whether
a target or targets are present.  The  conditional distributions of measurements,
 $p(\x | T=0)$ and $p(\x | L, T=1)$, are Gaussian with all means equal to 0, and covariances $\C = \Ss + \Sx$ and $\tC = \mathbf{\Sigma }_{\mathbf{s}, L}^0 + \Sx$, respectively.
We therefore have
 (see  Appendix~\ref{appendix_multipleTST_eq.d1} for details)
\begin{equation}\label{eq:multipleTST_d1}
\dTD(\x) = \log{\left[ \Nchoosen^{-1} \sqrt{\frac{|\C|}{|\tC|} }\displaystyle{\sum_{L \in \cL}} \exp{\left(-\frac{1}{2}\x^T\left(\tC^{-1}-\C^{-1}\right)\x\right)}\right]}.
\end{equation}

This decision variable depends on the model parameters, and the measurement, $\x$. The total number of stimuli, $N,$ number of targets, $n,$ and the variability,  $\sigma_s^2$. The correlation, $\rs$, between the distractors determine the structure of the stimulus, while the variability, $\sigma_x^2$, and correlation, $\rx$, describe the distribution of sensory measurements.  An ideal observer knows all these parameters.

Setting the right hand side of Eq.~\eqref{eq:multipleTST_d1} to 0 defines a nonlinear decision boundary in the space of measurements.  
This boundary separates measurements that lead to a ``target present'' decision ($\dTD(\x) > 0$) from those that
lead to a ``target absent'' decision ($\dTD(\x) < 0$). For example, with a single stimulus, $N = n = 1$, the observer needs to decide whether or not a single measurement corresponds to a target. The decision variable has the
form 
$$
\dTD(\x) = \log \left( \sqrt{\frac{\sigma_s^2+ \sigma_x^2}{\sigma_x^2}} \right) -   \frac{1}{2 \sigma_x^2} 
\frac{x_1^2}{1+\sigma_x^2/\sigma_s^2},
$$
If the measurement, $x_1$, differs sufficiently from 0, then
$\dTD(\x) < 0$.  A measurement close to 0, gives $\dTD(\x) > 0$.

With more stimuli, the observer needs to take into account  the known correlations between the stimuli and between
the measurements.  
The decision variable  depends in a complicated way on the 
parameters that describe the structure of the stimulus and the response.  The explicit 
form of Eq.~\eqref{eq:multipleTST_d1}, can be derived under the assumption of equal variances and equal covariances (See Appendix~\ref{appendix_multipleTST_eq.d1})
\begin{align}
\label{eq:multipleTST_d2}
& \dTD(\x) = \log \left( \frac1M \sqrt{ \frac{1+(\rs \sigs + \rx \sigx)Nv}{\gamk}\left(\frac{\tvj}{v}\right)^n } \sum_{L \in \cL} \exp{\left\{ -\frac{1}{2} \left( \underbrace{\sigma_s^2(1-\rho_s)\tvj v \sum_{i \in L}x_i^2 }_{\text{I}} \right. \right. }  \right. \nonumber \\
& \left. \left. \left. + \underbrace{\left(\beta v^2 -\frac{\rx \sigx \; \tvj^2 g_L}{\gamk} \right)\sum_{i,j \in L}x_ix_j }_{\text{II}} + 
\underbrace{2 \left(\beta v^2 - \frac{\rx \sigx \; \tvj v}{\gamk} \right)\sum_{i \in L, j \notin L}x_ix_j}_{\text{III}} + 
\underbrace{\left(\beta v^2 - \frac{v^2 f_L}{\gamk} \right)\sum_{i,j \notin L}x_ix_j}_{\text{IV}} \right) \right\} \right).
\end{align}

The variables $v$ and $\tvj$ represent scaled inverse variances corresponding to distractor and target stimuli. The parameters $\beta, g_L, f_L$, and $\gamk$ are given in Eqs.~\eqref{eq:useful_quantities_1} and~\eqref{eq:useful_quantities_2}, and are defined in terms of $\sigs, \rs, \sigx$, and $\rx$.  Eq.~\eqref{eq:multipleTST_d2} has a form that can be interpreted intuitively: 
 \begin{enumerate}
 \item Term I contains a sum of squares of individual measurements, $x_i^2,$ over the putative set of targets, $L$. 
 The smaller this sum, the more likely that $L$ contains targets.
\item  Term II
contains the sample covariance about the known target value, $s_T = 0$, that is,  $\sum_{i,j \in L} (x_i - s_T)(x_j -s _T) =  \sum_{i,j \in L} x_i x_j $. In the absence of measurement correlations, $\rx = 0$, covariability about the target value has vanishing expectation for target measurements.  Therefore the larger this sum, the less likely that the measurements come from a  set of target stimuli.  In the presence of measurement correlations, $\rx >0$, covariability between target measurements  is expected. Hence the prefactor in term II decreases with $\rx$.
\item Term III is similar to term II, with the sum representing the  sample covariance about the target value between putative target and non-target stimuli.  The larger this covariance, the less likely that $L$ or its complement contain targets.
\item Term IV contains the sample covariance about the mean of measurements outside of the putative target set. 
If there are no targets, then all terms in the sum are expected to be large, regardless of the choice of $L$.  
However, if there are targets,
then whenever  the complement of $L$ contains targets, some of the terms in the sum have expectation 0.
Hence the term again makes a smaller contribution if targets are present. 
 \end{enumerate}
 
While this provides an intuitive interpretation of the sums in Eq.~\eqref{eq:multipleTST_d2}, the expression is complex and it is difficult to precisely understand how an ideal observer uses 
knowledge of the generative model and the stimulus measurements to make a decision.    We therefore examine a
number of cases where Eq.~\eqref{eq:multipleTST_d2} is tractable, and all the terms 
can be interpreted precisely.  We also numerically examine performance in a wider range of examples.

\subsection{Single target, $\boldsymbol{n = 1}$ }

We start with the case when a single target is present at one of $N$ locations. This case was considered previously in the absence of correlations between the sensory measurements~\citep{Bhardwaj2015}.

\begin{figure}[htp!]
\begin{center}
\includegraphics[scale = 0.88]{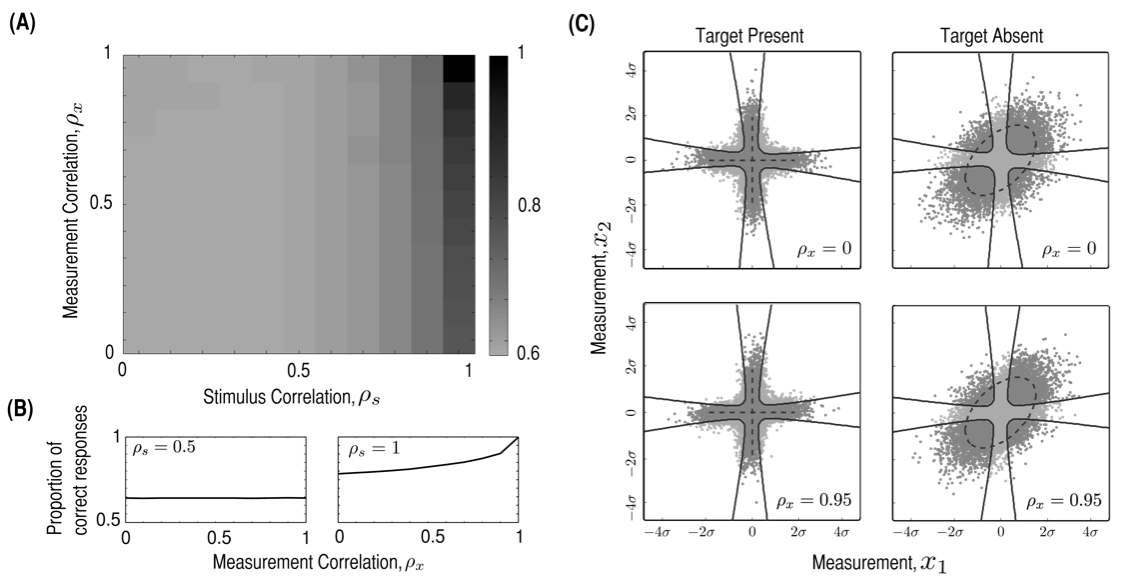}
\caption{Performance of an optimal observer on a single target detection task with weak measurement noise, $\sigx \ll \sigs$. \textbf{(A)} Proportion of correct responses as a function of stimulus correlation, $\rs$ and measurement correlation, $\rx$, for $N = 4$ stimuli. \textbf{(B)} Proportion of correct responses as a function of measurement correlation, $\rx,$ in the case of intermediate stimulus correlation, $\rs = 0.5$ (left) and strong correlation, $\rs = 1$ (right) for $N = 4$ stimuli. \textbf{(C)} Decision boundary, $\dTD(\x) = 0$ (black solid line) and measurement distributions on target present (left) and target absent (right) trials for $N = 2$ and $\rs = 0.5$. Here and henceforth dark gray dots correspond to measurements that result in a correct inference, and light gray dots correspond to measurements leading to an incorrect inference. The stimulus will result in a measurement that lies within the black dashed lines with probability 0.95. Other parameters used: $\sigma_s = 15$ and $\sigma_x = 4$.}
\label{fig:singleTST_perf_weak}
\end{center}
\end{figure}

\begin{figure}[htp!]
\begin{center}
\includegraphics[scale = 0.88]{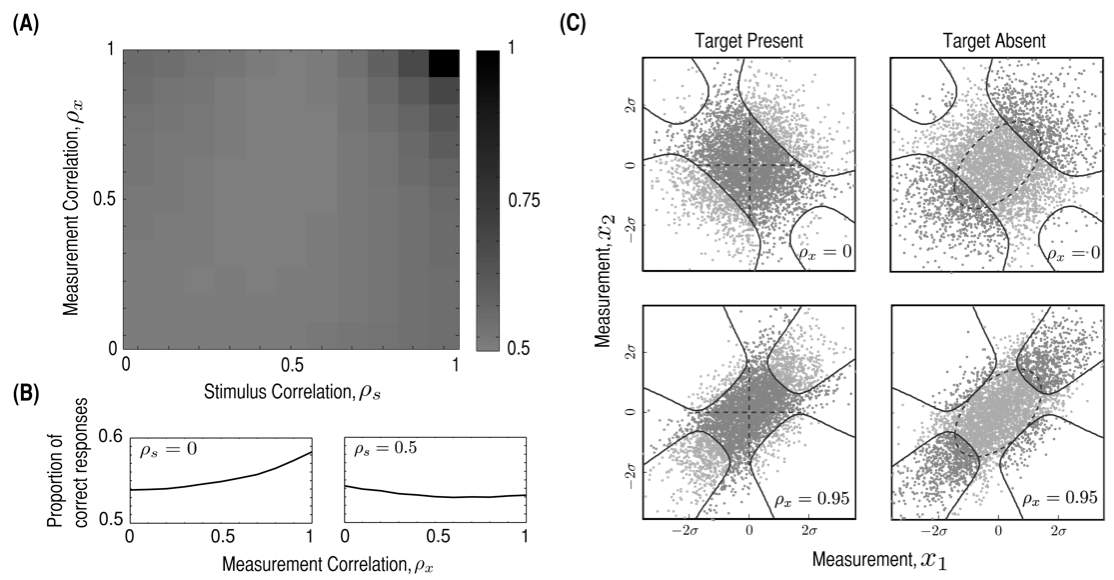}
\caption{Performance of an optimal observer on a single target detection task with strong measurement noise, $\sigx = \sigs$. \textbf{(A)} Proportion of correct responses as a function of $\rs$ and $\rx$ for $N=4$ stimuli. \textbf{(B)} Performance as a function of measurement correlation, $\rx,$ when $\rs = 0$ (left) and $\rs = 0.5$ (right) for $N=4$ stimuli.  (\textbf{C}) Decision boundary (black solid line) and distribution of measurements, $\x$ on target present (left) and target absent (right) trials for $N=2$ and $\rs = 0.5$. Other parameters used: $\sigma_s = \sigma_x = 15$.} 
\label{fig:singleTST_perf_strong}
\end{center}
\end{figure}

We observe in Figs.~\ref{fig:singleTST_perf_weak}A and~\ref{fig:singleTST_perf_strong}A that the performance of an ideal observer  is nearly independent of $\rs$ and $\rx$ when external structure  is weak ($\rs < 1$).  Performance 
depends strongly on $\rx$  when distractors are strongly correlated,  $\rs \approx 1$. An ideal observer performs perfectly when $\rs = \rx = 1$ (Fig.~\ref{fig:singleTST_perf_weak}C).  

Increased performance with increasing stimulus correlations, $\rho_s$,  accords with intuition that similar distractors make 
it easier to detect a target. However, correlations in measurement noise can play an equally important role and significantly improve performance when distractors are identical (See Fig.~\ref{fig:singleTST_perf_weak}B).  

Perfect performance when $\rs = \rx = 1$, can be understood intuitively.  In this case measurements, $x_i$, of the stimuli are obtained by adding the \emph{same} realization of a random variable, \emph{i.e.} identical measurement noise to each stimulus value, $s_i$. In target absent trials, all measurements are hence identical.   If the target is present, measurements contain a single outlier.  An ideal observer can thus distinguish the two cases perfectly.

We examine in more detail the cases of weak measurement noise, and then the case of comparable measurement noise and distractor variability.  
 
\paragraph{Weak measurement noise, $\boldsymbol{\sigx \ll \sigs}$ with highly correlated stimuli, $\rs \approx 1$.}  
When measurement noise is weak, discriminability between the ``target absent'' and ``target present'' conditions
is governed by external variability, \emph{i.e.} trial-to-trial variability of the stimuli.  As noted,  measurement correlations improve discriminability in the presence of strong stimulus correlations (see Figs.~\ref{fig:singleTST_perf_weak}A,C). When $\rs \approx 1$, Eq.~\eqref{eq:multipleTST_d2} can be approximated as (details in Appendix~\ref{appendix_singleTST_weakintnoise_rs1}):
\begin{align}\label{eq:singleTST_d_weaknoise_rs=1}
\dTD(\x) \approx & \log{\left(\underset{P(N,\rx)}{\underbrace{{\frac1N} \sqrt{ \frac{N(1-\rx)}{N-1} }}} \right.}  \nonumber \\
& \left. \sum_{i = 1}^N \underset{\alpha_i(\x,N,\rx, \sigma_x)}{\underbrace{\exp{\left \{-\frac{1}{2N\sigx(1-\rx)}\left((1- N\rx)x_i^2 + 2 x_i \sum_{j\neq i} x_j - \frac{1}{N-1} \left(\sum_{j \neq i} x_j\right)^2\right)\right \}}}}\right).
\end{align}
The ideal observer hence uses knowledge about the measurement correlations, $\rx$, in a decision. 

We first confirm the intuitive  observation that an ideal observer performs perfectly 
when measurement noise is highly correlated, $1 - \rx  \ll 1$.  In this case the exponential term in  Eq.~\eqref{eq:singleTST_d_weaknoise_rs=1} is, to leading order in $1/(1 - \rx)$,
\begin{align} \label{E:exp}
\alpha_i(\x,N,\rx,\sigma_x) \approx 
\exp{\left[\frac{N-1}{2N\sigx(1 - \rx)}\left(x_i - \bar{x}_{\hat{\imath}} \right)^2\right]},
\end{align}
where $\bar{x}_{\hat{\imath}} = 1/(N-1)\sum_{j \neq i}x_j$ is the sample mean of the measurements \emph{excluding} the putative target, $i$.
Hence, to make a decision, the ideal observer subtracts the mean of the $N-1$ measurements of putative distractors from  that of the putative target.  
In Appedix~\ref{appendix_singleTST_weakintnoise_rs1} we show that on ``target absent'' trials, $\dTD(\x) \rightarrow  -\infty$,  as $\rx \to 1$, and on target present trials, $\dTD(\x) \rightarrow  \infty$, as $\rx \to 1$. Hence performance improves as measurement correlations increase.  This can also be seen in Fig.~\ref{F:overlap}B, as the overlap between the distributions  $p(\x | T=0)$ and $p(\x | T=1)$ decreases with an increase
in $\rx$.

When measurement correlations are absent, $\rx = 0$, an ideal observer
 performs perfectly only when measurement noise is vanishing.  The exponential in Eq.~\eqref{eq:singleTST_d_weaknoise_rs=1} now  equals
\begin{align} \label{E:exp2}
\alpha_i(\x,N,\rx,\sigma_x) &= \exp{\left[-\frac{1}{2\sigx}\left(\frac{1}{N}\left(\sum_{j} x_j\right)^2 
- \frac{1}{N-1}\left(\sum_{j \neq i}x_j \right)^2
\right)\right]} \notag \\
&=  \exp{\left[-\frac{1}{2\sigx}\left(  N \bar{x}^2 - (N-1) \bar{x}^2_{\hat{\imath}}
\right)\right]},
\end{align}
where $\bar{x}$ is the sample mean of all observations.  The ideal observer therefore compares the sample mean over \emph{all measurements}, $\bar{x}$, with the sample mean over all observations excluding the putative target, $\bar{x}_{\hat{\imath}}$.  Since measurement noise is uncorrelated, averaging over all measurements is beneficial. This is a very different strategy than when measurement noise is highly correlated (Eq.~\eqref{E:exp}), and the ideal observer compares a \emph{single}
putative target measurement, $x_i$, to the sample mean of the  remaining target measurements.
As shown in Appendix~\ref{appendix_singleTST_weakintnoise_rs1} at intermediate values of measurement correlations, $0 < \rx < 1$, the ideal observer
uses a mixture of these two strategies.

Intuitively, an increased number of distractors make
it more difficult to detect the target.   An analysis of the decision variable in the limit $N \rightarrow \infty$ shows
that this is indeed the case (See Appendix~\ref{S:increasingN}).

%

\paragraph{
Weak measurement noise,  non-identical distractors, $\rs < 1$.}  
Measurement correlations have little effect on performance when stimulus correlations are weaker (See Fig.~\ref{fig:singleTST_perf_weak}A).    Consider again $\rx \approx 1$, so that measurements are corrupted by adding a random, but nearly identical perturbation to the stimuli. An ideal observer uses the knowledge that all 
measurements are obtained by adding an approximately equal value to the stimulus.
However, when stimulus correlations are weak, the target measurement is no longer an outlier.  Correlations in measurement noise  provide little help in this situation.

These observations are reflected in the structure of the decision boundaries ($\dTD(\x) = 0$), and the distributions of the measurements (See Fig.~\ref{fig:singleTST_perf_weak}B).  In the target absent (left column) and present trials  (right column) the
distributions of measurements, $p(\x |T)$, is shaped predominantly by variability in the stimulus. 
Measurement correlations have little effect on this shape, and the decision boundary therefore changes little
with an increase in $\rx$.   In contrast when $\rs \approx 1$, measurement correlations significantly impact the overlap between the distributions $p(\x | T = 1)$ and $p(\x | T = 0)$ as shown in  Fig.~\ref{F:overlap}B.

We can confirm this  intuition about the role of measurement correlations by approximating the decision 
criterion when $\sigx \ll \sigs$ and $\rs < 1$ (see details in Appendix~\ref{appendix_singleTST_weakintnoise_rs<1}), 
\begin{equation}\label{eq:singleTST_d_weaknoise_rs<1}
\dTD(\x) \approx \log{\left(\frac{1}{N} \sqrt{ \frac{\sigs (1-\rs)(1+(N-1)\rs)}{\sigx(1+(N-2)\rs)} }\sum_{i = 1}^N \exp{\left \{-\frac{x_i^2}{2\sigx}\right \} }\right)}. 
\end{equation}
The strength of  noise correlations, $\rx,$ does not affect  the decisions of an ideal observer at highest order in $(\sigx/\sigs)$.  This explains the approximate independence of performance on $\rx$ observed in Fig.~\ref{fig:singleTST_perf_weak}B. 
In this case an ideal observer considers stimuli at each location  separately, and 
weighs each measurement individually by its precision, $1/\sigx$~\citep{Green:1966,Palmer:1999}. 

When the number of distractors becomes larger, Eq.~\eqref{eq:singleTST_d_weaknoise_rs<1} is no longer valid. 
The observer compares measurements of all stimuli to make a decision.  We return to this point below.

\paragraph{Strong measurement noise, $\boldsymbol{\sigx = \sigs}$ } 
 Increasing measurement noise trivially degrades performance.  However, in the limit of perfect stimulus and measurement correlations, an ideal observer still performs perfectly for the  reasons described earlier. 

Measurement correlations affect performance differently than in the case of weak measurement noise (See  Fig. \ref{fig:singleTST_perf_strong}A,C).  Even with uncorrelated stimuli, $\rs = 0$, performance increases slightly (approximately 5-6\%) with $\rx$. Surprisingly, for intermediate values of stimulus correlations, \emph{e.g.} $\rs = 0.5$, measurement correlations negatively impact performance.  If measurement correlations are fixed at a high value, then the worst performance is observed at an intermediate value, $0 < \rs < 1$.   The  reason for this unexpected behavior is unclear, as Eq.~\eqref{eq:multipleTST_d2} is difficult to analyze in this case. 

Generally, when measurement noise is strong, measurement correlations  will change the shape of the 
measurement distributions $p({\bf x} |  T = 0)$ and $p({\bf x} |  T = 1),$
and hence
 impact decisions and performance.
Note that when measurement correlations increase,  the region corresponding to $\dTD(\x) > 0$ is elongated along the diagonal to capture more of the mass of the distribution $p({\bf x} |  T = 1)$ (See Fig.~\ref{fig:singleTST_perf_strong}C).  
However, when measurement noise is high, the interactions between measurement and stimulus correlations are intricate.

\subsection{Multiple targets, $\boldsymbol{n>1}$}

In the present task when multiple targets are present, they are all identical  and hence perfectly correlated. Thus, regardless of the value of $\rs$, 
on half the trials the stimuli will be strongly structured, and the density $p(\s | T = 1)$  concentrated on a low dimensional subspace. As a consequence, measurement correlations always  impact performance.  

\begin{figure}[htp!]
\begin{center}
\includegraphics[scale = 0.88]{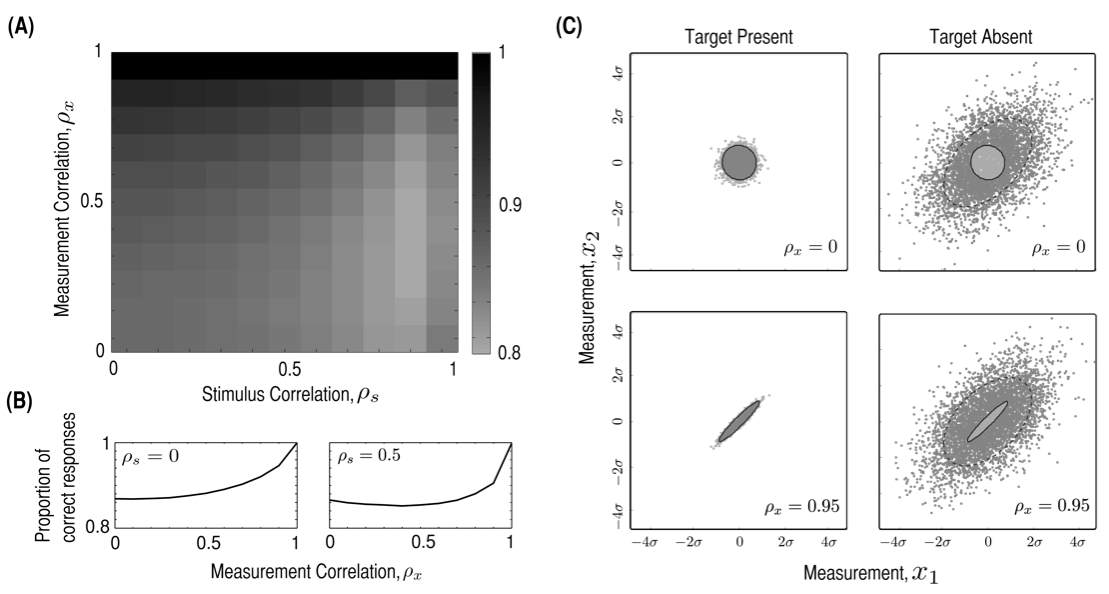}
\caption{ Performance of an optimal observer in a multiple target search task with weak measurement noise, $\sigx \ll \sigs.$
\textbf{(A)} Proportion of correct responses as a function of stimulus correlation, $\rs,$ and measurement correlation, $\rx,$ with $N = 4$ stimuli and $n = 3$ targets. \textbf{(B)} Proportion of correct responses as a function of $\rx$ for $\rs = 0$ (left) and $\rs = 0.5$ (right) when $N=4$ and $n=3$.  (\textbf{C}) Measurement distributions and decision boundary on target present (left) and target absent (right) trials for $\rs = 0.5$ and $N=n=2$.}
\label{fig:multipleTST_perf_weak}
\end{center}
\end{figure}

Regardless of stimulus correlations, an ideal observer performs perfectly when $\rx = 1$ (See Fig.~\ref{fig:multipleTST_perf_weak}A).   Even when $\rs < 1$, performance increases with $\rx$ (see Fig.~\ref{fig:multipleTST_perf_weak}A). When 
$\rx = 1$,  all target measurements are identical. Hence, an ideal observer performs perfectly 
by checking whether $n$ of the measurements, $x_i,$  are equal.
We only analyze the case $\rs < 1$, since the case of perfectly correlated distractors
is similar.

%

With weak measurement noise, Eq.~\eqref{eq:multipleTST_d2} can be approximated as (see Appendix~\ref{appendix_singleTST_weakintnoise_rs<1}):
\begin{align}\label{eq:multipleTST_d_weaknoise_rs<1}
\dTD(\x) \approx & \log\bigg(\frac{1}{M} \sqrt{ \frac{(1-\rx)(1+(N-1)\rs)}{(1+(N-n-1)\rs)(1+(n-1)\rs)}\left(\frac{\sigs(1-\rs)}{\sigx(1-\rx)}\right)^n } \nonumber \\ 
& \times \sum_{L \in \cL} \exp{\left\{-\frac{n}{2\sigx(1-\rx)} 
\underline{\left( \frac1n \sum_{i\in L}x_i^2 - \frac{n \rx}{(1+ (n-1) \rx)}\left(\frac{1}{n} \sum_{i \in L}x_i\right)^2 \right) } \right \}  } \bigg).
\end{align}
An ideal observer takes into account measurement correlations for all values of $\rs$ even when measurement noise is low.  
Stimulus correlations only appear  in the prefactor, and are not used in the comparison of the measurements inside the exponential.

Interestingly, decisions in this case are based only on measurements of stimuli within the set of putative targets, $L$. In the absence of  measurement correlations, a decision is based solely on the sample second moment of the $n$ stimulus measurements \emph{about the target characteristic, $s_T = 0$}, \emph{i.e.} $ 1/n \sum_{i\in L}x_i^2$ (the underlined term in Eq.~\eqref{eq:multipleTST_d_weaknoise_rs<1}).  A low value of this sample moment indicates that $L$ contains targets.   

In the limit of $\rx \to 1$, the underlined term in Eq.~\eqref{eq:multipleTST_d_weaknoise_rs<1} approaches the sample
variance, $\text{s}^2_{i\in L}$, \emph{i.e.} the sample second moment about the sample mean,  
\begin{equation}\label{eq:explimit}
\frac{1}{n}\sum_{i\in L}x_i^2 - \frac{n\rx}{1+\rx(n-1)}\left(\frac{1}{n}\sum_{i \in L}x_i\right)^2 \to \frac{1}{n}\sum_{i \in L}x_i^2 - \left(\frac{1}{n}\sum_{i \in L}x_i\right)^2 = \text{s}^2_{i\in L}.
\end{equation}
If measurement correlations are strong, the measurements of a set
of target stimuli will be approximately equal (regardless of $\rs$). The ideal observer makes use of this knowledge by comparing the value of putative targets in the set $L$.  If the sample variance of the measurements $\{x_i\}_{i \in L}$ is small, then it is likely that $L$ is a set of targets. When $\rx = 1$, an ideal observer performs perfectly (See Appendix~\ref{S:perfect} for details). 

When $0 < \rx < 1$, the underlined term in Eq.~\eqref{eq:multipleTST_d_weaknoise_rs<1} shows that the ideal observer takes an intermediate strategy by computing a 
second moment about a point between the target characteristic, $s_T$,  and the sample mean.  
Interestingly, the larger the number of targets, the larger the weight on the sample mean,
since the prefactor $n \rx /(1+ (n-1) \rx)$ increases with $n$ for fixed $\rx$.

These observations are reflected in the distributions shown in Fig.~\ref{fig:multipleTST_perf_weak}C.   The distribution of measurements, $p({\bf x} |  T = 1)$, move closer to the diagonal ($x_1=x_2$) as $\rho_x \to 1$, and the overlap with the distribution of measurements, $p({\bf x} |  T = 0)$ decreases.   In higher dimensions, for $N$ stimuli and $n$ targets, the measurement distributions, $p({\bf x} |  T = 1)$, is concentrated on the union of $(N-n+1)$-dimensional subspaces when $\rx =1$: The target measurements lie on a line, while the $N-n$ distractor measurements are distributed along the remaining directions.  

To conclude, when there are multiple targets part of the stimulus set is always perfectly correlated.  When measurement correlations are high,
the observer checks whether the measurements are similar to each other to make a decision. When measurement
correlations are low, the observer compares the measurements to the known target value.  
Measurement correlations can again decrease the overlap between
the conditional distributions of measurements, and significantly impact decisions and performance.  
For finite $N$, decisions are based on the comparison of measurements within a putative set of stimuli, $L$. 
We show next that when $N$ is large, this is no longer the case.

\subsection{Larger number of targets and stimuli}

If we assume a fixed proportion, $K$, of the stimuli consists of targets, so that
$n = KN$, then Eq.~\eqref{eq:multipleTST_d2} simplifies considerably in the limit of large $N$.
If we let $c_x = \sigx(1-\rx)$ and $c_s = \sigs (1-\rs)$, the exponential in 
Eq.~\eqref{eq:multipleTST_d2} has the form, (See Appendix~\ref{analysis:largeN_n})
\begin{equation} \label{E:largeN}
\alpha_L(\x,N,c_x,c_s) \approx \exp \left( -\frac12  \left[
   \underbrace{ \frac{NK}{c_x} s^2_{i \in L}}_{I} - 
    \underbrace{ \frac{N}{c_s + c_x}   s^2}_{II}  +  \underbrace{ \frac{N(1-K)}{c_s + c_x}  s_{i \not \in L}^2}_{III}
\right] \right), 
\end{equation}
where $s^2_{i \in L}$, $s_{i \not \in L}^2$, and $s^2$ are the sample variances of measurements form the 
putative target set, $L$, outside of the putative target set, and over all $N$ measurements, respectively.

The different terms in this expression can be interpreted as earlier:
Term I is the sample variance of measurements of the putative targets.  If this variance is large, then
the set $L$ is unlikely to contain targets. Term II is the sample variance among all terms.  If this term
is large then all stimuli are dissimilar, and  there is evidence that targets are present:
For example, when distractors are correlated, and $c_s \ll 1$, then the sample variance of all stimulus measurements is small
only in the absence of targets.  Finally, term III is the sample variance among putative distractors.  If distractors
are correlated, this term will be small if $L$ contains targets, and hence stimuli outside $L$ are distractors.   The sign of the three terms agrees with
this interpretation:  Terms I and III are negative, and term II is positive.

The main difference between the cases of large $N$, and the examples discussed previously is that 
an observer takes into account putative distractor measurements, \emph{i.e} measurements outside the
putative target set $L$.  An exception is the case when measurement correlations are much stronger 
than distractor correlations,
$c_x \ll c_s$.  In this case, the putative targets are more strongly structured, and hence only their
measurements are used in a decision.  
When $c_s \ll 1$, or, equivalently, $\rs \approx 1$, and distractors are strongly
correlated, all three terms in Eq.~\eqref{E:largeN} are comparable.  In this case, ideal observers
base their decision on the similarity, as measured by sample variance, of both putative distractor
and target measurements.  

\begin{figure}[htp!]
\begin{center}
\includegraphics[scale = 0.7]{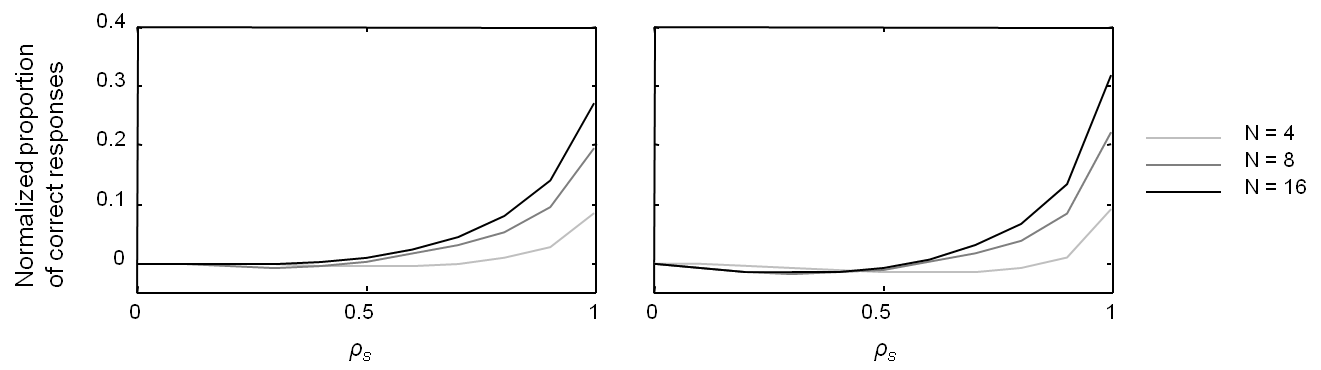}
\caption{Normalized performance of an optimal observer with respect to $\rs = 0$ as a function of stimulus correlation, $\rs$, for different values of $N$ in case of (A) $\rx = 0$ and (B) $\rx = 0.5$. 
Intermediate correlations between distractor measurements become increasingly important as
the number of distractors increases from four to sixteen.
Other parameters used: $\sigma_s = 15$, $\sigma_x = 4$, and $n = 2$.
} 
\label{fig:pc_diffn}
\end{center}
\end{figure}

Importantly,  the decision is made using distractor measurements, even 
when distractors are not perfectly correlated.  Fig~\ref{fig:pc_diffn} shows that 
intermediate distractor correlations have an increasingly impact decisions 
with an increase in distractor number. Indeed, the higher the fraction of distractors, $(1-K)$, the 
more weight is assigned to their sample variance (term III).
This is unlike the case of small $N$, where distractor
measurements are used only when they are perfectly correlated.

%
%

\section{Discussion}

We have shown that in a simple task the statistical structure of the stimulus as well
as that of  noise in  perceptual measurements determine the strategy and performance of 
an ideal observer.  Correlations in measurement noise can significantly impact performance,
particularly when stimulus correlations are high: When the distribution of stimuli 
conditioned on the parameter of interest is 
concentrated in a small volume of stimulus space, the statistical structure of
measurement noise can be of particular importance   
~\citep{Mazyar2012, Mazyar2013, Bhardwaj2015}.

The impact of noise correlations on the inference of 
a parameter from neural responses has been studied in detail~\citep{Averbeck2006, Averbeck2009, Latham2005, Perkel1967, Schneidman2003, Sompolinsky2001}.   Frequently the parameter of interest was identified with the stimulus, and both were
 univariate. The estimation of the orientation of a bar in the receptive field of a population
of neurons  has been a canonical example. 

Reality is far more complex.  Stimuli, such as a natural visual or auditory scene, are high dimensional and highly structured. 
Moreover, only some of the parameters are typically relevant. Intuitively, if noise perturbs measurements along relevant direction, \emph{i.e.} along the directions of the parameters of interest, then estimates will be corrupted. Perturbations along irrelevant directions in parameter space have little effect~\citep{MorenoBote:2014}. Measurement correlations can channel noise into irrelevant directions, without decreasing overall noise magnitude, and thus improve parameter inference.

This is difficult to study using general theoretical models without putting some constraints on the 
structure of measurement noise.  We therefore considered a relatively simple, analytically 
tractable example where both measurement and stimulus structure are characterized by a small number
of parameters. We have used a similar setup to examine decision making in controlled search experiments~\citep{Mazyar2012, Mazyar2013, Bhardwaj2015}.

We assumed that  measurement noise and measurement correlations can be varied independently. 
This is not realistic. For instance, it is known that changes in the mean, variability and covariability of neural responses can be tightly linked~\citep{Cohen:2011, delaRocha2007, Rosenbaum2011}. It is thus likely that the statistics of measurement noise also change in concert.  However, at present this relationship has not been well characterized. 

More importantly, noise from the periphery
of the nervous system will limit the performance of any observer.
It is therefore not possible that a simple change in measurement 
correlations can lead to perfect performance~\citep{MorenoBote:2014}.
To address this question it would be necessary to provide a more accurate model
of both the noise correlations in a recurrent network encoding information about the
stimuli~\citep{Beck:2011}, as well as the resulting measurement correlations.  
This is beyond the scope of the present study.

We also made strong assumptions about the structure of measurement and stimulus correlations. We chose to restrict our analysis 
to positive correlations.  The reason is that the requirement that a covariance matrix is 
positive definite implies restrictions on the range of allowable negative measurement and stimulus
correlations~\citep{horn2012}.  These restrictions depend on the number of stimuli, $N$, and complicate the analysis.
To make the model tractable, we also assumed that all off-diagonal elements in the stimulus and measurement noise covariance
matrices are identical.  While we did not examine it here, heterogeneity in the correlation structure can strongly affect parameter inference~\citep{Shamir:2006,Chelaru:2008,Berens:2011}. 

\begin{figure}[htp!]
\begin{center}
\includegraphics[scale = 0.75]{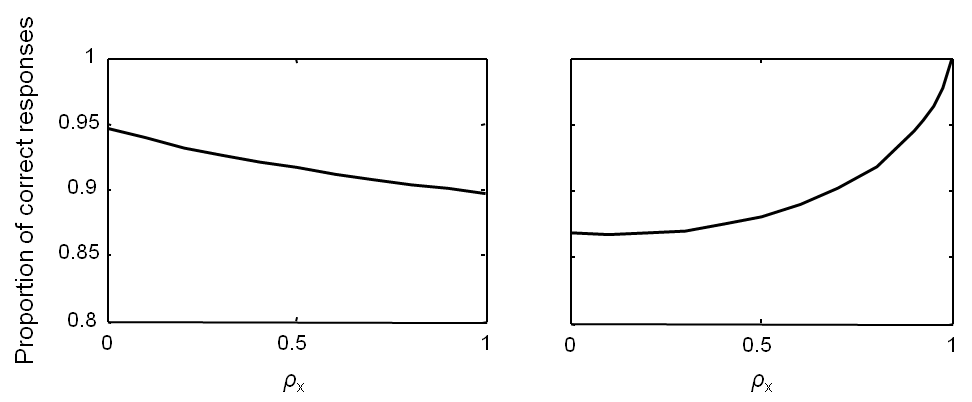}
\caption{Performance comparison of an optimal observer as a function of measurement correlations in  \textbf{(A)}  a mean left/right discrimination task with $N = 4$ stimuli and \textbf{(B)}  a target detection task with $n = 3$ targets. Other parameters used: $\sigma_s = 15$, $\sigma_x = 4$, and $\rs = 0.5$.}
\label{fig:intropanel}
\end{center}
\end{figure} 

To end, we provide another illustration of the fact that measurement correlations can affect the performance of
an ideal observer in different ways depending on the task:
Suppose an observer is presented with $N$ oriented stimuli, such as Gabor patches. The  stimuli and measurements follow the same Gaussian distributions introduced earlier in this study. The observer is asked to perform one of the following two tasks: 1) Report whether the mean orientation of the stimuli is to the left or right of vertical; 2) Report whether a vertically oriented target is present or absent. In this case, a subset of the stimuli has vertical orientation on half the trials. 

The first task is a discrimination task, and the observer needs to to integrate information from different sources.  
When measurement correlations are high, it is more difficult to average out the noise between the stimulus measurements~\citep{Sompolinsky2001,Zohary:1994ei}. The estimate of the average orientation is therefore degraded and performance
decreases with an increase in measurement correlations (see ~Fig.~\ref{fig:intropanel}A and Appendix \ref{appendix_meanleft/right}).
The second is a detection task that requires extracting information that is
 buried in a sea of distractors.   As discussed above, in this case measurement correlations can increase performance 
 if there is more than one target, or if the distractors are strongly correlated (see ~Fig.~\ref{fig:intropanel}B).  
 
In this example the stimuli and the measurements have the same statistical structure on target absent trials for the 
two tasks. However, the parameter of interest differs:  In the first task the observer needs to estimate the 
average stimulus orientation, and in the second determine whether a target is present.  The distributions
of measurements conditioned on these parameters are therefore also different, and are differently affected
by measurement noise.  

The question of how measurement correlations impact decision making and performance 
 does not have a simple answer~\citep{Hu:2014}.  Correlations in measurement noise can have a pronounced effect when 
the stimuli themselves are highly correlated, \emph{i.e.} when they occupy a small
volume in stimulus space.  We have illustrated how in this case measurement correlations 
can help in separating the distribution of measurements conditioned on a parameter of interest.
Similar considerations will be
important whenever we try to understand how information can be extracted from 
the collective responses of neural populations to high dimensional, and highly structured stimuli.

\section{Acknowledgment}
K.J. was supported by NSF award DMS-1122094. W.J.M. was supported by award number R01EY020958 from the National Eye Institute and award number W911NF-12-1-0262 from the Army Research Office.

\begin{appendices}

\section{Derivation of  Eqs.~\eqref{eq:multipleTST_d1} and~\eqref{eq:multipleTST_d2} }
\label{appendix_multipleTST_eq.d1}

Here we present the details of some of the calculations leading to the results presented in 
the main text.  The computations rely on the assumption of Gaussianity.  We will
therefore make repeated use of the fact that the density $\N(\x; \boldsymbol{\mu}, \boldsymbol \Sigma)$ of
the normal distribution with mean $\boldsymbol{\mu}$, and covariance $\boldsymbol \Sigma$
is 
\begin{equation}
\N (\x; \boldsymbol{\mu}, \boldsymbol \Sigma) = \frac1{\sqrt{(2 \pi)^{N} | \boldsymbol{\Sigma} |}} \exp{ 
\left(-\frac12 (\x - \boldsymbol \mu)^\text{T} \boldsymbol \Sigma^{-1} (\x - \boldsymbol \mu) \right)},
\end{equation}
where $| \boldsymbol{\Sigma} |$ denotes the determinant of the matrix $\boldsymbol{\Sigma}$.

Let $M = \Nchoosen$ denote the cardinality of the set $\cL$ of all possible sets $L$.
We compute $p(\x | T = 1)$  in Eq.~\eqref{eq:decision_TD} by marginalizing over $\s$,
$$
p(\x | T = 1) = \int p(\x | \s) p(\s | T = 1) d\s.
$$
We note that $$p(\s |T = 1) = \displaystyle{\sum_{L\in \cL} } p(\s | T= 1, L) p(L) = \frac1M \; \displaystyle{\sum_{L \in \cL} } p(\s | T= 1, L).$$ 
Therefore,
\begin{align*}
p(\x | T = 1) & = \frac1M \int p(\x | \s) \displaystyle{\sum_{L \in \cL} } p(\s | T= 1, L)  d\s \\
& = \frac1M \; \displaystyle{\sum_{L \in \cL} } \int p(\x | \s) p(\s | T= 1, L)  d\s \\
& = \frac1M \; \lim_{\eta \to 0} \displaystyle{\sum_{L \in \cL} } \int \N(\x; \s, \Sx) \N(\s; \0_N, \tS) d\s \\ 
& = \frac1M \; \displaystyle{\sum_{L \in \cL} } \N(\x; \0_N, \tC),
\end{align*}
Similarly,
\begin{align*}
p(\x | T = 0) & = \int p(\x | \s) p(\s | T= 0)  d\s \\
& = \int \N(\x; \s, \Sx) \N(\s; \0_N, \Ss) d\s \\  
& = \N(\x; \0_N, \C),
\end{align*}
where  $\tC = \Sx + \mathbf{\Sigma }_{\mathbf{s}, L}^0$ and $\C = \Sx + \Ss$.
We therefore obtain 
\begin{align}\label{eq:multipleTST_limd1}
\dTD(\x) & = \log \frac{p(\x | T = 1)}{p(\x | T = 0)} = \log \left(\frac1M\; \displaystyle{\sum_{L \in \cL}} \frac{\N(\x; \0_N, \tC)}{\N(\x; \0_N, \C)}  \right) \nonumber \\
& = \log \left( \frac1M \; \sqrt{\frac{|\C|}{|\tC|}} \displaystyle{\sum_{L \in \cL} } \exp{\left(-\frac{1}{2}\x^T\left(\tC^{-1}-\C^{-1}\right)\x \right)}  \right).
\end{align}
We note that the determinant of $\tC$ does not depend on the set $L$ since all matrices $\tC$ can be obtained from each other by permuting appropriate rows and columns.

In the case variances and covariances are equal, we can invert $\tC$ and $\C$. In general, matrix $\tC$ has the following form,
\begin{eqnarray*}
(\tC)_{i,j} = 
\begin{cases}
\sigx, & \text{ if } i = j \in L, \\
\sigs + \sigx, & \text{ if } i = j \notin L, \\
\rx \sigx, & \text{ if } i \neq j, \text{ and } i \text{ or } j \in L, \\
\rs \sigs + \rx \sigx, & \text{ if } i \neq j, \text{ and } i,j \notin L, \\
\end{cases}
\end{eqnarray*} 

We will use the Sherman-Morrison-Woodbury formula~\citep{meyer2000}, 
\begin{equation} \label{E:SMW}
\left(A + UEV \right)^{-1} = A^{-1} - A^{-1}U \left(E^{-1}+VA^{-1}U \right)^{-1} VA^{-1}, 
\end{equation}
to obtain the inverses of $\tC$ and $\C$. To do so we first rewrite $\tC$ as $A_L + U_L E V_L$: When  $L = \{1,2,...,n\}$ (targets placed at the first $n$ out of $N$ possible locations), we have
\begin{align*}
A_L & = \begin{bmatrix}
(\sigx - b) I_n & 0_{n \times (N-n)} \\
0_{(N-n)\times n} & (\sigs + \sigx - a) I_{(N-n)}
\end{bmatrix}_{(N \times N)},
\qquad 
& U_L  & = \begin{bmatrix}
b & \cdots & b & a & \cdots & a \\
b & \cdots & \cdots & \cdots & \cdots & b
\end{bmatrix}_{(N \times 2)}^{\text{T}}, \\
V_L & = \begin{bmatrix}
0 & \cdots & 0 & 1 & \cdots & 1\\
1 & \cdots & 1 & 0 & \cdots & 0
\end{bmatrix}_{(2 \times N)},
\quad \text{and} \quad
& E & = \begin{bmatrix}
1 & 0 \\ 0 & 1
\end{bmatrix}_{(2 \times 2)},
\end{align*}
where $a = \rs \sigs + \rx \sigx$, and $b = \rx \sigx$.

Using Eq.~\eqref{E:SMW} we obtain
\begin{eqnarray*}
(\tC^{-1})_{i,j} = 
\begin{cases}
\tvj - \frac{b \tvj^2 g_L}{\gamk}, & \text{ if } i = j \in L, \\
v - \frac{v^2 f_L}\gamk , & \text{ if } i = j \notin L, \\
-\frac{b \tvj^2 g_L}\gamk, & \text{ if } i \neq j, \text{ and } i, j \in L, \\
-\frac{v^2 f_L}\gamk, & \text{ if } i \neq j, \text{ and } i,j \notin L, \\
-\frac{b v \tvj}\gamk, & \text{ if } i \neq j, i \in L, j \notin L \text{ or } i \neq j, i \notin L, j \in L, \\
\end{cases}
\end{eqnarray*} 
where
\begin{align}\label{eq:useful_quantities_1}
v & = \frac1{\sigs ( 1- \rs) +\sigx (1 - \rx)}, \quad \tvj = \frac1{\sigx (1 - \rx)},  \quad f_L = a + n \rs \sigs \rx \sigx \; \tvj  \nonumber \\
g_L & =1 + \rs \sigs \; (N-n)v, \quad \text{and} \quad \gamk = 1+ a(N-n)v + n \rx \sigx \; \tvj g_L. 
\end{align}
Using the Matrix Determinant Lemma~\citep{Harville1998}, we can obtain the determinant of $\tC$
$$
\text{det}(\tC) = \frac{\gamk}{\tvj^n \; v^{N-n}}.
$$

Similarly, we compute the inverse and determinant of matrix $\C$,
\begin{equation*}
\C^{-1} = \begin{bmatrix}
v-\beta v^2 & -\beta v^2 & \cdots & -\beta v^2 \\
-\beta v^2 & v-\beta v^2 & \cdots & -\beta v^2 \\
\vdots & & \vdots & \\
-\beta v^2 & -\beta v^2 & \cdots & v-\beta v^2
\end{bmatrix},
\end{equation*}
where
\begin{equation}\label{eq:useful_quantities_2}
\beta = \frac{a}{1+ a N v}, 
\end{equation}
and
$$
\text{det}(\C) = \frac{1+a N v}{v^N}.
$$
The prefactor in Eq.~\eqref{eq:multipleTST_limd1}, is therefore
\begin{equation*}
\sqrt{\frac{|\C|}{|\tC|}} = \sqrt{\frac{1+aNv}{\gamk}\left(\frac{\tvj}{v}\right)^n }
\end{equation*}
and we can compute
\begin{eqnarray*}
\x^T\C^{-1}\x & = & (v-\beta v^2)\sum_{i=1}^{N}x_i^2 - \beta v^2\sum_{i \ne j}^Nx_ix_j, \\
\x^T\tC^{-1}\x & = & \left( \tvj -\frac{1}{\gamk}b\tvj^2 g_L \right)\sum_{i \in L}x_i^2 + \left( v - \frac1\gamk v^2 f_L \right)\sum_{i \notin L}x_i^2 -\frac{1}{\gamma_k}b\tvj^2g_L \sum_{\substack{i,j \in L \\ i \ne j}}x_ix_j \\
& & -\frac2\gamk b v \tvj \sum_{\substack{i \in L \\ j \notin L}}x_ix_j - \frac{1}{\gamma_k}v^2f_L\sum_{\substack{i,j \notin L \\ i\ne j}}x_ix_j.
\end{eqnarray*}
A slight rearrangement of the terms therefore shows that  when variances and covariances are equal, Eq.~\eqref{eq:multipleTST_limd1} is equivalent to
\begin{align*}
&\dTD(\x)  = \log \left( \frac1M \sqrt{ \frac{1+a N v}{\gamk}\left(\frac{\tvj}{v}\right)^n } \sum_{L \in \cL} \exp{\left\{-\frac{1}{2} \left( \sigma_s^2(1-\rho_s) \tvj v\sum_{i \in L}x_i^2 \right. \right. }  \right. \nonumber \\
& \left. \left. \left. + \left(\beta v^2 -\frac{\tvj^2 b g_L}{\gamk} \right)\sum_{i,j \in L}x_ix_j  + 2 \left(\beta v^2 - \frac{\tvj v b}{\gamk}\right)\sum_{i \in L, j \notin L}x_ix_j + \left(\beta v^2-\frac{f_L v^2}{\gamma_L} \right)\sum_{i,j \notin L}x_ix_j \right) \right\} \right).
\end{align*}

\section{Asymptotic analysis of Eq.~\eqref{eq:multipleTST_d2} }
\label{appendix_approximations}
Here we present some asymptotic results for the decision variable $\dTD (\x)$ given in Eq.~\eqref{eq:multipleTST_d2}. The main results are obtained for small measurement noise, $\sigx$. 
Equivalent results can be obtained for large external variability, $\sigs$.

\subsection{Small measurement noise and idental distractors,  $\boldsymbol{\rs = 1}$} 
\label{appendix_singleTST_weakintnoise_rs1}

We first concentrate on the case $n = 1$.  The exponential terms in Eq.~\eqref{eq:multipleTST_d2} simplify to the following:
\begin{align*}
\sigs (1-\rs)\tvj v & = 0, \\
\beta v^2 - \frac{\rx \sigx \tvj^2 g_L}{\gamk} & = \frac{\sigs +\rx \sigx}{\sigx(1-\rx)[N \sigs+ \sigx(1 + (N-1) \rx)]} \\
& - \frac{\rx [\sigs(N-1) + \sigx(1-\rx)]}{\sigx(1-\rx)[\sigs (N-1) + \sigx(1-\rx)(1+(N-1)\rx)]} \\
& = \frac{1-N \rx}{N \sigx(1-\rx)} + \mathcal{O}(1),  \\
\beta v^2 - \frac{\rx \sigx \tvj v}{\gamk} & = \frac{\sigs + \sigx\rx}{\sigx(1-\rx)[N \sigs+ \sigx(1 + (N-1) \rx)]} - \frac{\rx}{\sigs(N-1) + \sigx(1-\rx)(1+(N-1)\rx)} \\
& = \frac{1}{N\sigx(1-\rx)}+\mathcal{O}(1), \\
\beta v^2 - \frac{v^2 f_L}{\gamk} & =  \frac{\sigs + \sigx\rx}{\sigx(1-\rx)[N \sigs+ \sigx(1 + (N-1) \rx)]} \\
& - \frac{\sigs + \rx \sigx (1-\rx)}{\sigx(1-\rx)[\sigs(N-1) + \sigx(1-\rx)(1+(N-1)\rx)]} \\
& = -\frac{1}{N(N-1)\sigx(1-\rx)}+\mathcal{O}(1).
\end{align*}
And the leading determinant term becomes:
\begin{align*}
\sqrt{\frac{1+a N v}{\gamk}\frac{\tvj}{v}} & = \sqrt{ \frac{(1-\rx)[N \sigs + \sigx(1+ (N-1) \rx)]}{\sigs(N-1) + \sigx(1-\rx)(1+ (N-1) \rx)}} \\
& = \sqrt{\frac{N(1-\rx)}{N-1} + \mathcal{O}\left(\sigx \right)}
\end{align*}
Therefore, the decision variable becomes approximately,
\begin{align*}
\dTD(\x) \approx & \log{\left(\frac{1}{N} \sqrt{\frac{N(1-\rx)}{N-1}} \right.} \nonumber \\
& \left. \sum_{i = 1}^N \exp{\left \{\frac{-1}{2N\sigx(1-\rx)}\left((1- N\rx)x_i^2 + 2 x_i \sum_{j\neq i} x_j - \frac{1}{N-1} \left(\sum_{j \neq i} x_i\right)^2\right)\right \} }\right).
\end{align*}
After rearranging terms, this can be rewritten as 
\begin{align}\label{eq:mixed}
\dTD(\x) \approx & \log \left(\frac{1}{N} \sqrt{\frac{N(1-\rx)}{N-1}}  \right. \nonumber \\
& \left. \sum_{i = 1}^N \exp{\left[ \rx \; \frac{N-1}{2N\sigx (1 - \rx)}\left(x_i - \bar{x}_{\hat{\imath}} \right)^2
-\frac{1}{2\sigx}\left(  N \bar{x}^2 - (N-1) \bar{x}^2_{\hat{\imath}} \right) \right]  } \right),
\end{align}
where  $\bar{x}$ is the sample mean of all measurements, and where $\bar{x}_{\hat{\imath}}$ is the sample mean of the measurements \emph{excluding} the putative target, $i$.  In the limiting cases $\rx = 0$ and $(1 - \rx) \ll 1$, we obtain the 
exponents given in Eq.~\eqref{E:exp} and Eq.~\eqref{E:exp2} discussed in the text. 

\subsection{Perfect performance when $\rx = 1$}

We show that when $\rs = 1$, an ideal observer performs perfectly in the limit of identical measurement noise. For a fixed number of stimuli, $N,$ on ``target absent'' trials, $x_i - (N-1)^{-1}\sum_{j \neq i}x_j = O(\sqrt{\epsilon})$, and hence $\alpha(\x,N,\rx,\sigma_x) = O(1)$. On the other hand the prefactor,
$P(N,\rx)=O(\sqrt{\epsilon})$, and hence $\dTD(\x) \rightarrow  -\infty$, as $\epsilon \rightarrow 0$, \emph{i.e.} as $\rx \to 1$. 
On ``target present'' trials, when stimulus $i$ is the target, then $x_i - (N-1)^{-1}\sum_{j \neq i}x_j = O(1)$, and 
$\alpha_i(\x,N,\rx,\sigma_x) = O(1/\epsilon)$.   In this case the prefactor is still $P(N,\rx)=O(\sqrt{\epsilon})$, and the exponential 
term dominates. A similar argument works for the summands for which $i$ is not a target. 

\subsection{Single target with increasing number of distractors}
\label{S:increasingN}

We still work under the assumption that
measurement noise is relatively weak, so that we can use Eq.~\eqref{eq:mixed}.  Note that on ``target absent'' trials, 
$\bar{x} = s + O(\sigx / \sqrt{N})$, where $s$ is the true value  of the (identical) distractors.  We also have 
$\bar{x}_{\hat{\imath}} = s + O(\sigx / \sqrt{N-1})$.   Hence, the first term in the exponential of
Eq.~\eqref{eq:mixed} is $O(1)$, while the second term is
$O(\sqrt{N})$.  A similar argument holds in ``taget present'' trials.

On target absent trials
$$
-\frac{1}{2\sigx}\left(  N \bar{x}^2 - (N-1) \bar{x}^2_{\hat{\imath}} \right) = \exp{\left[-\frac{1}{2\sigx} s^2 + O(\sqrt{N} \sigma_x) \right]} 
\qquad \text{for all } i,
$$
to leading order in $N$. We abuse notation slightly and only use order notation on
the terms that include measurement noise.
As  stimuli become more 
dissimilar to the target, \emph{i.e.} as $s^2$ increases, $\alpha_i$ decreases
exponentially,  $\dTD(\x)$ becomes more negative, and it is hence easier to infer that a target is absent.  However, the
$O(\sqrt{N} \sigma_x)$ terms can be both positive and negative.  Thus performance 
decreases with the number of stimuli.  
Similarly we can see that when a target is present
\begin{align} 
\alpha_i(\x,N,\sigx) &= \exp{\left[\frac{1}{2\sigx} \frac{N-1}{N} s^2 + O(\sqrt{N} \sigma_x) \right]} 
\qquad \text{when  $i$ is the target}, \label{E:first} \\
\alpha_i(\x,N,\rx) &= \exp{\left[- \frac{1}{2\sigx} \frac{1+N - N^2}{N - N^2} s^2 + O(\sqrt{N} \sigma_x) \right]} 
\qquad \text{when  $i$ is not a target}, \label{E:second}
 \end{align}
again to leading order in $N$.   As measurement noise decreases, or $s^2$ increases, the first term given in Eq.~\eqref{E:first} diverges exponentially, and the terms given in Eq.~\eqref{E:second} approach 0 exponentially.  As a result $\dTD(\x)$ increases.
However, the $O(\sqrt{N} \sigma_x)$ noise term increases with $N$,  
and an increase in
the number of stimuli again decreases performance.

If $(1 - \rx) \ll 1/N$, \emph{i.e.} measurement noise is strongly correlated, the first term in the exponential of
Eq.~\eqref{eq:mixed} dominates.  Thus when correlations increase faster than the inverse
of the number of distractors, performance increases with 
the number of distractors.

\subsection{Weak external structure, $\boldsymbol{\rs < 1}$, arbitrary number of targets}
\label{appendix_singleTST_weakintnoise_rs<1}

We approximate each term in the exponential of Eq.~\eqref{eq:multipleTST_d2} assuming $\sigx \ll 1$:
\begin{align*}
\sigs (1- \rs) \tvj v & = \frac{\sigs (1-\rs)}{\sigx(1-\rx) [\sigs(1-\rs) + \sigx(1-\rx)] } \\
& = \frac{1}{\sigx(1-\rx)} + \mathcal{O}(1), \\
\beta v^2 - \frac{\rx \sigx \tvj^2 g_L}{\gamk} & = \frac{\rs \sigs +\rx \sigx}{[ \sigs(1-\rs) + \sigx(1-\rx)] [\sigs(1+(N-1)\rs) +\sigx(1+(N-1)\rx)]} \\
& - \frac{\rx [\sigs(1+(N-n-1) \rs + \sigx(1-\rx)]}{\sigx(1-\rx)[\sigs (1+(n-1) \rx) (1+(N-n-1)\rs) + \sigx(1-\rx)(1+(N-1)\rx))]}  \\
& = - \frac{\rx}{\sigx(1-\rx)(1+ (n-1)\rx)} + \mathcal{O}(1), \\
\beta v^2 - \frac{\rx \sigx \tvj v}{\gamk} & = \frac{\rs \sigs +\rx \sigx}{[ \sigs(1-\rs) + \sigx(1-\rx)] [\sigs(1+(N-1)\rs) +\sigx(1+(N-1)\rx)]} \\
& -\frac{\rx}{[\sigs(1+(N-n-1)\rs)(1+(n-1)\rx)+\sigx(1-\rx)(1+(N-1)\rx)]} \\
& = \mathcal{O}(1), \\
\end{align*}
and
\begin{align*}
& \beta v^2 - \frac{v^2 f_L}{\gamk} = \frac{\rs \sigs + \rx \sigx}{[\sigs(1-\rs) + \sigx(1-\rx)] [\sigs (1+(N-1)\rs)+\sigx(1+(N-1)\rx)]}  \\
& -\frac{\rs \sigs (1+(n-1)\rx) + \rx \sigx(1-\rx)}{[\sigs(1-\rs ) + \sigx(1-\rx)] [\sigs(1+(N-n-1)\rs)(1+(n-1)\rx)+\sigx(1-\rx)(1+(N-1)\rx)]} \\
& =  \mathcal{O}(1)
\end{align*}

We also approximate the leading coefficient of the exponential term in Eq.~\eqref{eq:multipleTST_d2} as:
\begin{align*}
& \sqrt{\frac{1+ (\rs \sigs + \rx \sigx) N v}{\gamk} \left(\frac \tvj v \right)^n } = \\
& \sqrt{\frac{(1-\rx)[\sigs(1+(N-1)\rs)+\sigx(1+(N-1)\rx)]}{[\sigs(1+(N-n-1)\rs)(1+(n-1)\rx) + \sigx(1-\rx)(1+(N-1)\rx)]}\left( \frac{\sigs (1-\rs) + \sigx(1-\rx)}{\sigx(1-\rx)} \right)^n} \\
& = \sqrt{ \frac{(1-\rx) ( 1+ (N-1)\rs)}{(1+(N-n-1)\rs) (1+(n-1)\rx)} \left( \frac{\rs(1-\rs)}{\sigx(1-\rx)} \right)^n +\mathcal{O}(1)}.
\end{align*}

Combining above terms, Eq.~\eqref{eq:multipleTST_d2} reduces to the following expression under the assumption of $\sigx \ll \sigs, \rs < 1$,
\begin{align*}
\dTD(\x) \approx & \log\bigg(\frac{1}{M} \sqrt{ \frac{(1-\rx)(1+(N-1)\rs)}{(1+(N-n-1)\rs)(1+(n-1)\rs)}\left(\frac{\sigs(1-\rs)}{\sigx(1-\rx)}\right)^n } \nonumber \\ 
& \times \sum_{L \in \cL} \exp{\left\{-\frac{1}{2\sigx(1-\rx)}\left(\sum_{i\in L}x_i^2 - \frac{\rx}{1+ (n-1) \rx}\sum_{i,j \in L}x_i x_j\right)\right \}  } \bigg).
\end{align*}

\paragraph{Special case: $\boldsymbol{n = 1}$} In case of a single target, set $L$ has only one element and $\cL = \{1, 2, \cdots, N \}$. In this case, Eq.~\eqref{eq:multipleTST_d_weaknoise_rs<1} reduces to a much simpler expression
\begin{equation*}
\dTD(\x) \approx \log{\left(\frac{1}{N} \sqrt{ \frac{\sigs (1-\rs)(1+(N-1)\rs)}{\sigx(1+(N-2)\rs)} }\sum_{i = 1}^N \exp{\left \{-\frac{x_i^2}{2\sigx}\right \} }\right)}.
\end{equation*}
This expression is independent of $\rx$. Hence, the decision boundary, and 
the performance of an ideal observer is unaffected by measurement correlations.

\subsection{Near perfect performance with $n > 1$, and $\rx \approx 1$}
\label{S:perfect}
We first assume that $T = 1$.  From Eq.~\eqref{eq:explimit}, if $L_T$ is the set of targets, then this expression is approximately zero, and the exponential is approximately unity.  When $L$ is a set not consisting of all targets, then the expression in Eq.~\eqref{eq:explimit} has expectation greater than zero.  Then 
\begin{eqnarray}
& & \sum_{L \in \cL}\exp\left\{-\frac{n}{2\sigx(1-\rx)}\left(\frac{1}{n}\sum_{i \in L}x_i^2-\left(\frac{1}{n}\sum_{i\in L}x_i\right)^2\right)\right\} = \nonumber \\
& & 1 + \sum_{L \in \cL \setminus {L_T}}\exp\left\{-\frac{n\alpha_L^2}{2\sigx(1-\rx)}\right\} \to 1
\end{eqnarray}
The prefactor in Eq.~\eqref{eq:multipleTST_d_weaknoise_rs<1}  diverges since the exponential dominates.  If  $T = 0$, then Eq.~\eqref{eq:explimit} will be greater than zero for all sets, $L$.  Thus $d_{TD}(\x) \to -\infty$ with $\rx \to 1$.

\subsection{Asymptotics for large $N$}

\label{analysis:largeN_n}

Here we develop asymptotic results for the decision variable when the number of stimuli, $N$ is large. 
We assume that there a fraction $K$ of the stimuli are targets, so that there are $NK$ targets
and $(1-K)N$ distractors. To simplify notation we write $c_x = \sigx(1-\rx)$ and $c_s = \sigs (1-\rs)$.
We find that 
\begin{eqnarray*}
(1-\rs)\tvj v & = & \frac{c_s}{c_x(c_s + c_x)} \\
v^2\left(\beta-\frac{1}{\gamk}\left(\frac{\tvj}{v}\right)^2\rx\sigx g_L \right) & = & \frac1N \frac{1}{c_s + c_x} - \frac{1}{NK} \frac{1}{c_x}  + O(\frac{1}{N^2})\\
v^2\left(\beta-\frac{1}{\gamk}\frac{\tvj}{v}\rx\sigx\right) & = & \frac1N \frac{1}{c_s + c_x} + O(\frac{1}{N^2}) \\
v^2\left(\beta-\frac{1}{\gamk}f_L \right) & = & \left(\frac1N - \frac{1}{(1-K)N} \right) \frac{1}{c_x + c_s}  + O(\frac{1}{N^2}) \\
\frac{1+ (\rs \sigs + \rx \sigx) N v}{\gamk} \left(\frac \tvj v \right)^n  &=& \left(1 + \frac{c_s}{c_x} \right)^{KN}  
\left(\frac1N \frac{(\rx - 1) (\rs \sigs + \rx \sigx)}{(K -1) K \rs \rx \sigs} \right)
\end{eqnarray*}
We can therefore group the terms in the exponential of the decision variable given by Eq.~\eqref{eq:multipleTST_d2}
as 
$$
-\frac{1}{2} \left( \frac{c_s}{c_x(c_s + c_x)} \sum_{i \in L}x_i^2 + \frac1N  \frac{1}{c_x + c_s} \sum_{i,j}^N x_ix_j 
- \frac{1}{NK} \frac{1}{c_x} \sum_{i,j \in L}x_ix_j - \frac{1}{(1-K)N}  \frac{1}{c_x + c_s}   \sum_{i,j \notin L}x_ix_j\right)
$$
A simple reorganization of the terms yields Eq.~\eqref{E:largeN}.

\section{Mean stimulus orientation - left or right discrimination task}
\label{appendix_meanleft/right}

An observer is presented with $N$ stimuli on every trial.  For concreteness, we can think of the stimuli as  bars, with orientations $\s = (s_1, s_2, \cdots, s_N)$.  The task is to decide whether the mean orientation of the set is to the left (a condition we denote by $C = -1$) or right ($C = 1$) of the vertical. The observer makes a decision based on the measurements, $\x = (x_1, x_2, \cdots,x_N)$. Stimulus orientations are drawn from a multivariate normal distribution with mean vector, $\0_N$, and covariance matrix, $\Ss$,  
\begin{equation}
p(\s) = \N(\0_N, \Ss), 
\end{equation}
with $\Ss$  defined in Eq.~\eqref{eq:matrix_Sigmas}. As in the target detection task, we assume the measurements  follow a multivariate normal distribution with mean vector and covariance matrix specified in Eq.~\eqref{eq:dist_x}.

An ideal observer performs the task by making a decision based on the log posterior ratio,
\begin{align} \label{eq:meantask_d1}
d_{\text{MOD}}(\x) & = \log \frac{p(C = 1| \x)}{p(C = -1| \x)}  = \log \frac{p(\bs >0 | \x)}{p(\bs < 0 | \x)} \nonumber \\
& = \log \frac{p(\x | \bs >0)}{p(\x | \bs < 0)} + \log \frac{p(\bs >0)}{p(\bs < 0)},
\end{align}  
where, $\bs = \displaystyle{\sum_{i = 1}^N s_i}$ denotes the mean stimulus orientation on a trial. If $d_{\text{MOD}}(\x) > 0$, the observer infers $\hC = 1$, that is, the mean stimulus orientation is to the right of the vertical. We compute Eq.~\eqref{eq:meantask_d1} by marginalizing over $\s$ and applying Bayes' rule,
\begin{eqnarray*}
p(\x|\bs >0) & = & \int p(\x|\s) p(\s|\bs >0)d\s \\
& = & \int p(\x|\s) p(\bs >0|\s)\frac{p(\s)}{p(\bs>0)}d\s \\
& = & \frac{1}{p(\bs > 0)}\int_{\bs > 0} p(\x|\s) p(\s)d\s \\
& = & \frac{1}{p(\bs > 0)}\int_{\bs > 0} \N(\x; \s, \Sx) \N(\s; \0_N, \Ss)d\s \\
& = & \frac{z_c}{p(\bs > 0)}\int_{\bs > 0} \N \left(\s; \left(\mathbf{I}+\Sx \Ss^{-1}\right)\x,\left(\Sx^{-1}+\Ss^{-1}\right)^{-1}\right)d\s,
\end{eqnarray*}
where $z_c$ is a normalization constant. Similarly, we compute $p(\x | \bs < 0)$ and obtain 
\begin{equation}\label{eq:decision_mean}
d_{\text{MOD}}(\x) = \log{\left(\frac{\int_{\bs > 0} \N \left(\s; \left(\mathbf{I}+\Sx\Ss^{-1}\right)^{-1}\x,\left(\Sx^{-1}+\Ss^{-1}\right)^{-1}\right)d\s}{\int_{\bs < 0} \N \left(\s; \left(\mathbf{I}+\Sx\Ss^{-1}\right)^{-1}\x,\left(\Sx^{-1}+\Ss^{-1}\right)^{-1}\right)d\s}\right)}.
\end{equation}

By symmetry, it is easy to see that the decision boundary in the space of measurements is given by 
the hyperplane $\bx = 0$, where  $\bx = \frac1N \displaystyle{\sum_{i = 1}^N x_i}$ is the sample mean of
the measurement.  
An ideal observer therefore bases the decision only on the sample mean. 
Therefore, we have $\bx > 0 \Rightarrow \bs > 0$. 

In order to understand the negative impact of noise correlations on the performance of an optimal observer on this task, consider $\x = \s + \vec{\xi}$ where $\vec{\xi} = (\xi_1, \xi_2, \cdots,\xi_N) \sim \N(0,\Sx)$, so that
$$
\bx = \frac1N \sum_i (s_i + \xi_i)  = \bs + \bar{\xi}.
$$
In this case, the variance of the mean noise scales with increasing noise correlation strength, $\rx$.
\begin{equation*}
\text{Var}(\bar{\xi}) = \frac{\sigma_x^2(1-\rho_x)}{N} + \rho_x\sigma_x^2.
\end{equation*}
As the variance increases, the overlap between the conditional distributions $p(\bx | \bs >0)$ and $p(\bx | \bs < 0)$ 
increases.  It is therefore more difficult to tell which condition the measurement comes from, and performance
deteriorates.

\end{appendices}

\nocite{*}
\bibliographystyle{apa}
\bibliography{references}
\end{document}